**All-optical Detection of Spin Hall Angle in W/CoFeB/SiO$_2$ Heterostructures by Varying Tungsten Layer Thickness**

Sucheta Mondal, Samiran Choudhury, Neha Jha, Arnab Ganguly, Jaivardhan Sinha and Anjan Barman[*]

Department of Condensed Matter Physics and Material Sciences,
S. N. Bose National Centre for Basic Sciences, Block JD, Sec. III, Salt Lake, Kolkata 700106 (India)
E-mail: (*abarman@bose.res.in)

The development of advanced spintronics devices hinges on the efficient generation and utilization of pure spin current. In materials with large spin-orbit coupling, the spin Hall effect may convert charge current to pure spin current and a large conversion efficiency, which is quantified by spin Hall angle (SHA), is desirable for the realization of miniaturized and energy efficient spintronic devices. Here, we report a giant SHA in beta-tungsten ($\beta$-W) thin films in Sub/W($t$)/Co$_{20}$Fe$_{60}$B$_{20}$(3 nm)/SiO$_2$(2 nm) heterostructures with variable W thickness. We employed an all-optical time-resolved magneto-optical Kerr effect microscope for an unambiguous determination of SHA using the principle of modulation of Gilbert damping of the adjacent ferromagnetic layer by the spin-orbit torque from the W layer. A non-monotonic variation of SHA with W layer thickness ($t$) is observed with a maximum of about 0.4 at about $t$ = 3 nm, followed by a sudden reduction to a very low value at $t$ = 6 nm. This variation of SHA with W-thickness correlates well with the thickness dependent structural phase transition and resistivity variation of W above the spin diffusion length of W, while below this length the interfacial electronic effect at W/CoFeB influences the estimation of SHA.

# I. INTRODUCTION

The exciting new frontier of spintronics [1] and magnonics [2, 3] research is driven by the need of utilizing spin-orbit (SO) effect for obtaining pure spin current [4]. An important aspect of improving the performance of the device is to minimize Joule heating, which requires fundamentally pure spin current [4, 5]. It is quite non-trivial to generate and transport the spin current. Some of the earlier studies have used non-local spin-valve based techniques [6, 7], spin pumping [8, 9, 10], and Rashba effect [11] for generating spin currents. Moreover, utilization of pure spin current for magnetization manipulation poses additional challenges. Recent finding of spin Hall effect (SHE) [12] has opened up the possibility of utilizing pure spin current for manipulation of magnetic moments [4, 13, 14]. To quantify the SHE, an important parameter, namely, spin Hall angle (SHA), has been proposed and it is related to the conversion efficiency of charge-to-spin current [15, 16]. Considerable efforts have been devoted for estimating and understanding the value of SHA for various heavy metals (HMs). Particularly, the SHE in HM layers can generate sufficiently large spin current to manipulate magnetic moments of a ferromagnetic layer adjacent to the HM layer as it exerts significant spin torque [13, 17]. Furthermore, the SHE induced spin-orbit torques (SOTs) have been shown to induce the large domain wall velocity [14], excite precessional magnetization dynamics [18, 19] as well as result in magnetization switching [20]. Remarkably, it has been recently demonstrated that by using sophisticated device structuring, SOT induced magnetization switching can be triggered in the absence of any magnetic field [21, 22, 23]. Some key requirements for technological implementation of the above mentioned interesting applications are to search for HMs with reasonably large SHA, investigation of various factors affecting the SHA of HM thin films, and to understand the variation of SHA in such HM thin films by controlling those factors. An important issue in this research is to establish an accurate and unambiguous measurement technique of SHA. The precise quantification of SHA and its origin in a conventional metal-based system is of technological interest for spintronics based device applications. In general, the techniques used for determining the SHA are the spin torque ferromagnetic resonance (ST-FMR) technique [13, 24, 25] spin torque switching of perpendicularly magnetized films [22] and measurement based on non-local spin valves [5]. All these techniques primarily rely on electrical excitation, detection and extremely delicate micro-fabrication [26]. Recently, it has been demonstrated that by using time resolved magneto-optical Kerr effect (TRMOKE) [27] technique SHA can be measured more conveniently in a non-invasive manner without the requirement of advanced micro-fabrication and electrical detection, and more precise estimate of SHA may be obtained [28].

The highly resistive $\beta$-tungsten (distorted tetragonal phase commonly referred to as A15 structure) is known to be one of the efficient materials for exhibiting large SHA due to strong SO coupling [29]. Also, in ferromagnetic thin film heterostructures, use of tungsten (W) leads to highly stable perpendicular magnetic anisotropy [30] and interfacial Dzyaloshinskii Moriya interaction [31]. Another important characteristics associated with W is the thickness dependent phase transition exhibited by it, usually observed in the thickness range of sub-10 nm [32, 33]. In general, sputter deposited W films with thickness below 5 nm is found to have $\beta$ phase with resistivity larger than 150 $\mu\Omega$-cm, whereas the films with thickness above 5 nm possess predominantly $\alpha$ phase (BCC structure) with resistivity of about 40 $\mu\Omega$-cm [20, 34, 35]. Till date, the SHA for W has been reported in few studies for mostly in $\beta$ phase. However, a systematic study of SHA in W/ferromagnet(FM)/oxide heterostructure with W layer thickness varying across the structural phase transition is missing. Depending on the deposition condition, SHA values of up to about 0.3 have been reported specifically for $\beta$ phase of W [20, 29]. Few reports have mentioned relatively small SHA for $\alpha$ phase of W [29]. Therefore, it calls for investigating the systematic dependence of SHA on the structural phase of W which is intricately related to its thickness. Furthermore, few important recent studies have suggested that the transparency of HM/FM interface plays a crucial role in evaluating the SHA of HM layer [36, 37, 38]. Additionally, a theoretical study has classified the bulk and interface SHE and claimed that the interface spin Hall effect may be as large as 25 times than the bulk SHE [39]. Recently, by alloying different HM layers, attempts have been made to achieve large SHA [40]. All these studies relate to the intricate role of spin orbit coupling induced SHE in generating pure spin current that is aimed towards utilization for device applications.

Here, we present the correlation between thickness dependent phase transition in W thin films and large SHE induced modulation of damping (MOD) in technologically important Sub/W($t$)/Co$_{20}$Fe$_{60}$B$_{20}$(3 nm)/SiO$_2$(2 nm) heterostructures. All-optical detection technique TRMOKE is used for investigating the magnetization dynamics [28]. Utilizing the sensitive variation of MOD, we estimate the SHA. We observe a clear variation in the value of estimated SHA with the phase of W underlayer. However, even within the $\beta$ phase, when the thickness of W is smaller than its spin diffusion length, the value of SHA is found to be significantly low. We correlate this variation of SHA with the bulk of the HM layer as well as the interfacial electronic effect at the HM/FM interface.

## II. SAMPLE PREPARATION AND CHARACTERIZATION

The thin film heterostructures Sub/W($t$)/Co$_{20}$Fe$_{60}$B$_{20}$(3 nm)/SiO$_2$(2 nm) with $t$ = 2 to 7 nm in the step of 1 nm were deposited by dc/rf magnetron sputtering on Si (100) wafers coated with 100 nm SiO$_2$. The purpose of varying W underlayer thickness was to choose W thickness across the phase transition regime. The base pressure of the deposition chamber was better than 2 × 10$^{-7}$ Torr. CoFeB and W were grown using dc power of 20 Watt whereas the SiO$_2$ was grown using rf power of 60 Watt at 13.56 MHz. All thin films were grown in Ar gas atmosphere of 1 mTorr pressure and deposition conditions were carefully optimized [31]. Using a shadow mask, 5 nm thick chromium/25 nm thick gold contact electrodes were first prepared, followed by deposition of the sample stack of 3 mm × 1 mm dimension between the contact electrodes using another shadow mask. The dc charge current was applied along the length of the sample using a standard source-meter [U3606A, Agilent Technologies] and experimental arrangement allowed us to suitably choose the applied bias magnetic field angle with respect to the current flow direction. Two-colour optical pump probe technique was used in the TRMOKE experiment. The second harmonic (wavelength: 400 nm, pulse width: 100 fs) of a mode locked Ti-sapphire oscillator (Tsunami, Spectra Physics) was used as the pump beam to excite the magnetization dynamics in the samples, whereas the fundamental laser beam (wavelength: 800 nm, pulse width: 80 fs) was used as the probe beam to detect the pump induced magneto-optical Kerr rotation from the sample as a function of the time delay between the pump and the probe beam [28]. All the experiments were performed under ambient condition. Atomic force microscope was used to investigate the surface topography, whereas vibrating sample magnetometer was used to characterize the static magnetic properties of these heterostructures. Using standard four probe technique the resistivity of W film was determined and the grazing incidence X-ray diffraction was used for investigating the phase of W.

### III. RESULTS AND DISCUSSION

Figure 1 (a) shows the grazing incidence X-ray diffraction (XRD) patterns for Sub/W($t$)/Co$_{20}$Fe$_{60}$B$_{20}$(3 nm)/SiO$_2$(2 nm). In these XRD plots, the peaks corresponding to $\alpha$ and $\beta$ phase of W are marked. The high intensity XRD peak at ~40.5° corresponds primarily to the $\alpha$ phase (BCC structure) of W (~40.5°) (110) orientation. Interestingly, we find that the peak in the vicinity of 40.5° is present for all thicknesses of W but when the W thickness is less than 6 nm, then the peaks (~34.8° and ~42.1°) corresponding to $\beta$-W (A15 structure) with (200) and (211) crystal orientations appear [34, 37]. One may note that in close proximity of 40°, $\beta$-W peak for (210) crystal orientation could also be present, which is quite difficult to identify. We wish to clarify here that for W thickness below 5 nm the $\beta$-rich phase along with small amount of $\alpha$ phase exists while with increasing thickness of the W layer the fraction of $\alpha$ phase increases and starts to dominate for W thickness above 5 nm. For the sake of simple notation we refer the phase below 5 nm W as *β*-phase and above this thickness as *α*-phase. These findings are consistent with some of the existing literature reports, where it is described that W exhibits a transition from *β* phase (A15 structure) to *α* phase (BCC structure) with increasing film thickness in the range of about 5 to 6 nm [32, 33]. It has also been shown in some other studies that this transition thickness may be increased or decreased by carefully tuning the deposition conditions of the W thin films [34, 41].

In Fig. 1 (b), the atomic force microscope images for all the heterostructures investigated in the present study are shown. From these images we observe that the average topographical roughness for the samples with $t$ = 2, 3, 4, 5, 6 and 7nmis 0.21, 0.21, 0.16, 0.19, 0.14 and 0.23 nm, respectively. The roughness values vary by about 10% when measured at various regions of space of the same sample. Overall, the topographical roughness in all film stacks is found to be significantly small irrespective of whether the W thickness corresponds to its *β* or *α* phase. Due to the small thicknesses of the thin film heterostructures, presumably, the interfacial roughness will clearly show its imprint on the topographical roughness. We thus infer that the interfacial roughness, if any, present in these heterostructures is very small and is similar in all samples. To determine the variation of resistivity of W with its thickness across the two different phases, we performed four probe measurements on all the samples. The inverse of sheet resistance of the film stack as a function of W thickness is plotted in Fig. 1(c). A change of the slope is observed beyond 5 nm, which indicates a change in the W resistivity. We estimate the average resistivity of W in *β* and *α* phase to be about 260 μΩ-cm and 105 μΩ-cm, respectively.

### A. Principle behind the determination of spin Hall angle

We next focus on the mechanism involved in the tuning of magnetization dynamics under the influence of spin current. Figure 2 (a) shows the schematic of experimental arrangement in which the flow of charge current through the W layer and consequent spin current generation due to SHE are shown. Under the influence of spin current, the CoFeB layer experiences an anti-damping like SOT and the magnetization dynamics is governed by modified Landau-Lifshitz-Gilbert equation [28] as given below:

$$\frac{d\hat{m}}{dt} = -\gamma(\hat{m} \times H_{eff}) + \alpha(\hat{m} \times \frac{d\hat{m}}{dt}) + \frac{\hbar}{2e\mu_0 M_s d} J_s (\hat{m} \times \hat{\sigma} \times \hat{m}) \quad (1).$$

Here, $\gamma$ is the gyromagnetic ratio, $\hat{\sigma}$ is the spin polarization vector, $\hat{m}$ is the magnetization vector, $M_s$ is saturation magnetization, $J_s$ is spin current density, $H_{eff}$ is the effective magnetic field, $d$ is the ferromagnetic layer thickness and $\alpha$ is the Gilbert damping constant [5, 42]. Depending on the polarity of $\hat{\sigma}$, the spin torque [24] acts collinearly against or towards the intrinsic Gilbert damping of the precessing magnetization. Effective damping, in turn, gets modulated depending on the injected spin current density and relative orientation between the magnetic moment (which lies along the direction of magnetic field) and charge current density [24]. The modulation of damping (MOD) under the influence of spin current [28, 43] can be expressed as:

$$\Delta\alpha = (\alpha - \alpha_0) = \hbar\gamma J_s / 2eM_s d 2\pi f \quad (2),$$

where $\alpha_0$ is the damping in the absence of applied charge current, $e$ is electronic charge, $f$ is the precessional frequency and other symbols have the same meaning as described before in the text. Thus, the SHA (charge current to spin current conversion efficiency) is given by:

$$\theta_{SH} = \frac{J_s}{J_c} = 2eM_s d 2\pi f \Delta\alpha / \hbar\gamma J_c \sin\theta \quad (3),$$

where, $\theta$ is the angle between $J_c$ and bias magnetic field direction. Experimentally, we find the value of $\Delta\alpha/J_c$ and estimate the SHA using the above expression.

**B. All-optical investigation of magnetization dynamics**

In Fig. 2(b), the as measured data using TRMOKE set-up is shown for the Sub/W(4 nm)/Co$_{20}$Fe$_{60}$B$_{20}$(3 nm)/SiO$_2$(2 nm) sample at $H$ = 1.46 kOe without any applied charge current. The femtosecond laser excites the sample thereby triggering the magnetization dynamics. Overall, the dynamics can be divided into three different temporal regimes as shown in the plot. Regime I (sharp drop immediately after negative delay, ~700 fs) corresponds to ultrafast demagnetization and regime II corresponds to the fast relaxation (1.7 ps) due to the spin lattice relaxation. Subsequently, in regime III, we observe a slower relaxation (~ 40 ps) along with magnetization precession, which gets damped in few ns. We mainly concentrate in regime III to estimate the damping and its modulation due to the action of spin torque. The blue line in Fig. 2 (b) corresponds to the bi-exponential background present in the precessional data in regime III. We subtract this background from the raw data and fit the resulting data using standard damped harmonic function. From the fit we estimate the damping $\alpha$ using the expression $\alpha = 1/2\pi f\tau$, where $f$ is the precessional frequency and $\tau$ is the relaxation time corresponding to magnetization oscillation.

We further studied the bias field dependent magnetization dynamics and Fig. 3(a) shows a representative experimental data of precessional magnetization dynamics along with the theoretical fit using a damped sinusoidal function for Sub/W(4 nm)/Co$_{20}$Fe$_{60}$B$_{20}$(3 nm)/SiO$_2$(2 nm) sample. From the fitting, we extract the relaxation time ($\tau$) as 0.52, 0.59 and 0.70 ns for bias field values of 1.46, 1.05 and 0.65 kOe, respectively. Figure 3 (b) shows the corresponding fast Fourier transform (FFT, power vs. frequency), from which the precessional mode frequency is extracted. The frequency ($f$) versus bias magnetic field ($H$) is plotted in Fig. 3 (c) for the same film stack. Standard Kittel expression mentioned below is used to fit the $f$ vs. $H$ data:

$$f = \frac{\gamma}{2\pi}\left( H\left(H + 4\pi M_{eff}\right)\right)^{\frac{1}{2}} \qquad (4),$$

where $\gamma = g\mu_B/\hbar$, $g$ is the Lande $g$ factor, $H$ is the applied bias magnetic field and $M_{eff}$ is the effective magnetization. From the fit $M_{eff}$ and $g$ are determined as fitting parameters. For these film stacks we obtain $M_{eff}$ ~ 1000 ± 30 emu/cc and $g$ = 2.0 ± 0.5, except for the W thickness of 5 nm, where $M_{eff}$ is found to be lower (770 ± 24 emu/cc). In Fig. 3(d), the $M_{eff}$ obtained from the dynamic measurement is plotted as a function of W thickness, $t$. Interestingly, for all the film stacks investigated in this study, $M_{eff}$ is found to be close to the saturation magnetization $M_s$ obtained using vibrating sample magnetometer. From this we infer that interface anisotropy is negligibly small in these heterostructures [31].

### C. Spin current induced modulation of damping

Figure 4 shows some typical time-resolved Kerr rotation data after the application of dc charge current through the film stack with different polarities and the MOD as a function of the dc charge current density. The applied charge current through the heterostructure gets distributed into W and CoFeB layers according to the resistivity of each metallic layer. Here, $J_c$ represents the current density through the W underlayer. Figure 4 (a) shows the magnetization precession data at a bias field $H$ = 1.46 kOe and for positive and negative $J_c$ ($\theta$ = 90°) along with the fit using damped sine function to extract the damping. The magnitude and sign of $J_c$ and the corresponding extracted value of effective damping $\alpha$ are mentioned in each panel. A direct comparison of MOD with applied $J_c$ is shown for three representative thickness of W in Fig. 4 (b) *i.e.*, $t$ = 3 nm, 4 nm (corresponding to $\beta$-rich phase) and 7 nm (corresponding to $\alpha$ phase). From these plots, it is evident that the spin current induced spin torque generated by SHE almost linearly modulates the damping consistent with the existing literature [24, 28]. We use a linear fit and extract the slope of MOD ($\Delta\alpha/J_c$) in order to estimate the SHA using Eq. 3 for a given thickness of W. The values of the slope extracted from the fit for 2 nm, 3 nm, 4 nm, 5 nm, 6 nm and 7 nm thick W are (0.47 ± 0.04) × 10$^{-12}$ m$^2$/A, (1.29 ± 0.10) × 10$^{-12}$ m$^2$/A, (1.20 ± 0.06) × 10$^{-12}$ m$^2$/A, (1.02 ± 0.04) × 10$^{-12}$ m$^2$/A, (0.39 ± 0.06) × 10$^{-12}$ m$^2$/A and (0.11 ± 0.01) × 10$^{-12}$ m$^2$/A, respectively. Note that the slope of MOD due to the spin current generated by SHE is generally larger for the $\beta$ phase W (3, 4, 5 nm) in comparison to the $\alpha$ phase W (6, 7 nm). The sample with 2 nm W thickness is an exception and it will be discussed later in this article. The limited $J_c$ value for the high resistive $\beta$ phase W (in Fig. 4 (b), for $t$ = 3 and 4 nm) is to avoid Joule heating in these films. Nevertheless, damping variation in $\alpha$ up to ±15% is observed for a reasonably small current density of $0.3 \times 10^{10}$ A/m$^2$ for the sample with $t = 3$ nm.

### D. Tungsten layer thickness dependence of spin Hall angle

Figure 5(a) shows the plot of SHA as a function of W thickness 2 nm $\leq t \leq$ 7 nm in Sub/W($t$)/Co$_{20}$Fe$_{60}$B$_{20}$(3 nm)/SiO$_2$(2 nm). Interestingly, in this plot we notice that the SHA is quite small when the W layer thickness is 2 nm, subsequently, SHA increases to a large value for W layer thickness of 3 nm and 4 nm. For $t>4$ nm, the SHA decreases monotonically up to $t = 7$ nm. It is important to emphasize here that we observe a giant value of SHA as large as 0.4 ± 0.04 for $t = 3$ nm, which is about 30% larger than the value of SHA reported using electrical detection technique (~0.3) [29]. Within the $\beta$ phase of W, the SHA decreases as the thickness of W becomes comparable to spin diffusion length of W ($\lambda_{sf}$) [41, 44]. The observed dependence of SHA on W thickness from 2 nm to 4 nm can be explained by considering drift diffusion analysis of the spin flow that incorporates spin Hall effect. Earlier, theoretical and experimental studies have proposed that within the spin diffusion length, the counter flowing spin current generated due to vertical gradient in the spin dependent electron chemical potential adjacent to the HM surface (under the assumption that no spin current penetrate out of HM) cancel the spin Hall generated spin current [20, 41, 43, 45, 46]. Due to this, the magnitude of the spin Hall spin current reduces significantly (resulting in smaller MOD and underestimation of SHA) as the thickness of the HM layer becomes comparable to $\lambda_{sf}$. From our experimental data we understand that the $\lambda_{sf}$ of our W thin film is less than 3 nm as it is difficult to extract this parameter precisely using standard fit with few numbers of data points available in $\beta$-W phase. In order to understand whether the variation of SHA is directly correlated with the variation of resistivity, we plot the W resistivity with its thickness in Fig. 5 (b). The significant drop in the resistivity value for $t > 5$ nm indicates a transition from $\beta$-W to $\alpha$-W phase in the W film used in our experiment. It is important to notice here that the variation in SHA above spin diffusion length is primarily correlated with the thickness dependent $\beta$ to $\alpha$ phase transition (structural change which is also related to the resistivity change c.f. Fig. 5(b)) of W [20, 29, 32, 34, 44]. Though from Fig. 5 (b), the resistivity of 5 nm thick W film is found to be primarily in $\beta$ phase, but from the trend of SHA values, it appears that there is probably a mixed $\beta$ and $\alpha$ phase of W at this thickness. Hence, apart from W thickness of 2 and 5 nm, SHA is found to be in direct correspondence with the resistivity, which is similar to the variation of SHA with conductivity as observed for other HMs [40, 47].

In the phase transition regime, the change in crystal structure indicates a change in SO coupling strength of W which may play an important role in modifying the SHA [32]. There is a possibility that in the low resistive regime (for $t > 5$ nm), the SHA is influenced by the change in SO coupling strength. As the phase transition in W with its thickness and the associated resistivity change predominantly originates from the bulk portion of W, thus, it indicates that the bulk part of W plays dominant role in defining the SHA and thereby the spin current above the spin diffusion length of W. The bulk SHA in the NM layers is theoretically predicted and experimentally found to consist of contributions from the intrinsic, the side-jump, and the skew scattering mechanisms [4, 48, 49]. The estimated large value of SHA in $\beta$-W films indicates that in these heterostructures all the mechanisms efficiently contribute for the large SHA as well as the W/CoFeB interface is highly transparent. The trend found in the variation of SHA with W thickness ($\beta$-W has larger SHA in comparison to $\alpha$-W) is mostly consistent with earlier reported results [29] while the values of SHA obtained using all-optical detection technique is a new addition in this field. The method employed here is non-invasive and more unambiguous as it helps to eliminate any experimental artifacts involved in the electrical detection schemes [50]. Being a local technique, the all-optical method does not suffer from the large area averaging, which could have produced spurious effects due to inhomogeneities and defects present in the sample. Further, in the time-domain measurement technique, magnetization damping can be directly extracted from time-resolved precession data, which is more advantageous than other techniques such as FMR line width measurement, where excitation of multiple modes may lead to inhomogeneous line broadening, which could artificially increase the damping.

## IV. CONCLUSION

In summary, we have used all-optical time-resolved magneto-optical Kerr effect microscopy to investigate the magnetization dynamics in Sub/W($t$)/Co$_{20}$Fe$_{60}$B$_{20}$(3 nm)/SiO$_2$(2 nm) with varying W layer thickness under the influence of spin current generated by SHE. The W layer thickness is so chosen that it undergoes a transition from $β$-rich to $α$-rich phase at a thickness above 5 nm. For highly resistive $β$ phase W, large modulation of damping of upto ±15% at a modest current density of $0.3 × 10^{10}$ A/m$^2$ and corresponding SHA as large as 0.4 is achieved. The SHA above the spin diffusion length of W follows the thickness dependent phase transition of W. On the other hand, a smaller modulation of damping and underestimation of SHA is observed for W layer thickness smaller than its spin diffusion length. In order to realize the full value of spin current due to SHE, it is thus important to use the HM thickness above its spin diffusion length. The variation of SHA with W thickness ($β$-W has larger SHA in comparison to $α$-W) is mostly consistent with the change in resistivity. Though in some of the recent theoretical and experimental studies, direct correspondence of SHA with conductivity has been discussed, however, we believe our investigations will trigger more studies to get further deep insight into the relationship between conductivity and spin Hall angle specifically for materials with mixed phase. Our results of detailed variation of SHA for different values of W thickness, will be beneficial for in-depth understanding of correlation between the thickness dependent phase transition in W and SHA. Furthermore, these studies will be significantly important from the application perspective as the future spintronic devices are expected to use large SHA material and spin current induced magnetization switching.

**Acknowledgements:** We gratefully acknowledge the support from S. Pal in sample characterization. We acknowledge the financial assistance from Department of Science and Technology Govt. of India under grant no. SR/NM/NS-09/2011 and S. N. Bose National Centre for Basic Sciences under project no. SNB/AB/12-13/96. SM acknowledges DST under INSPIRE scheme and SC acknowledges S. N. Bose National Centre for Basic Sciences for the senior research fellowship.

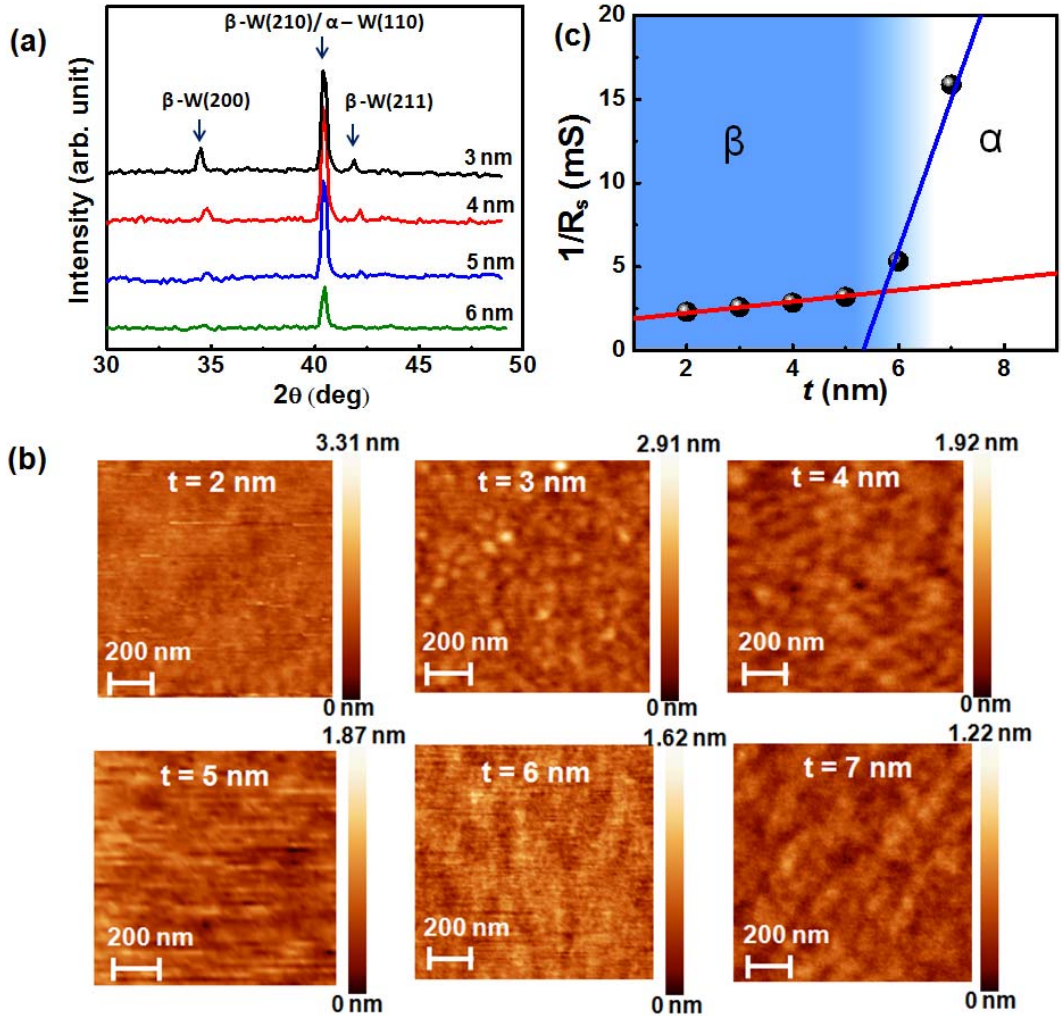

Figure 1. (a) X-ray diffraction patterns measured at grazing angle incidence for W films with thickness of 3 nm, 4 nm, 5 nm and 6 nm. Peaks corresponding to $\beta$ and $\alpha$ phase of W are marked in the plots. (b) Atomic force microscope images showing the surface topography of the Sub/W($t$)/Co$_{20}$Fe$_{60}$B$_{20}$(3 nm)/SiO$_2$(2 nm) samples with $t$ = 2 nm to 7 nm. (c) Variation of inverse of sheet resistance of Sub/W($t$)/Co$_{20}$Fe$_{60}$B$_{20}$(3 nm)/SiO$_2$(2 nm) as a function of W thickness ($t$) measured using linear four probe technique.

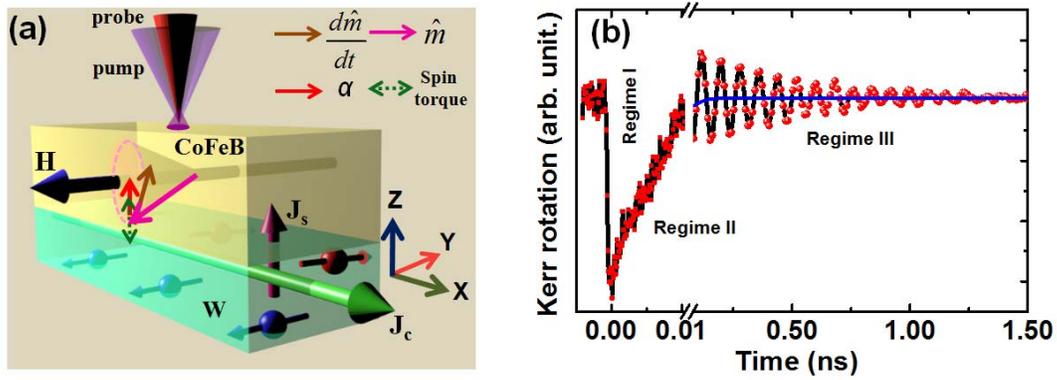

Figure 2. (a) Schematic of sample geometry and illustration of pump probe experimental geometry. Co-ordinate convention as followed is also shown. (b) Time-resolved Kerr rotation data for Sub/W(4 nm)/Co$_{20}$Fe$_{60}$B$_{20}$(3 nm)/SiO$_2$(2 nm) sample at H = 1.46 kOe is shown. The three different temporal regimes are indicated in the graph.

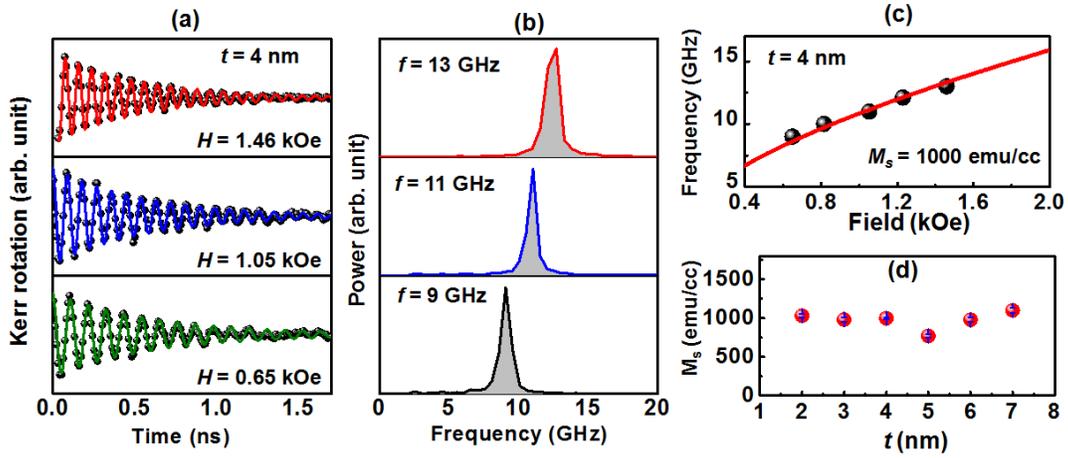

Figure 3. (a) Time-resolved precessional magnetization dynamics for Sub/W(4 nm)/Co$_{20}$Fe$_{60}$B$_{20}$(3 nm)/SiO$_2$(2 nm) sample at different bias magnetic field values. (b) The corresponding FFT power spectra to extract the precession frequency. (c) Plot of variation of frequency as a function of bias magnetic field. The solid line is the fit with Kittel formula. (d) Saturation magnetization of the Sub/W(*t*)/Co$_{20}$Fe$_{60}$B$_{20}$(3 nm)/SiO$_2$(2 nm) samples as a function of W layer thickness.

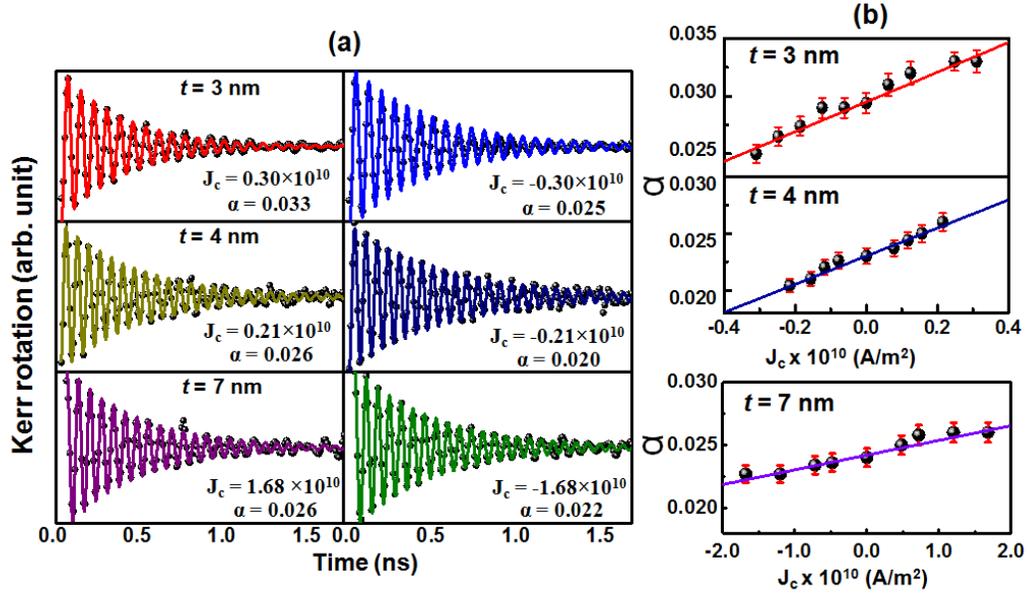

Figure 4. (a) Representative TRMOKE traces for extraction of damping under the influence of positive and negative current densities. Here, W thickness is mentioned in the left panel. The estimated damping values at mentioned current densities (in A/m$^2$) are also shown. Comparison of left and right panel indicates that the damping value changes with the polarity of charge current. (b) Modulation of damping plot for W thickness of 3 and 4 nm corresponding to the $\beta$ phase and 7 nm corresponding to the $\alpha$ phase. Solid line is the linear fit to the modulation of damping with current density. Error bars correspond to the fitting error obtained during the estimation of damping.

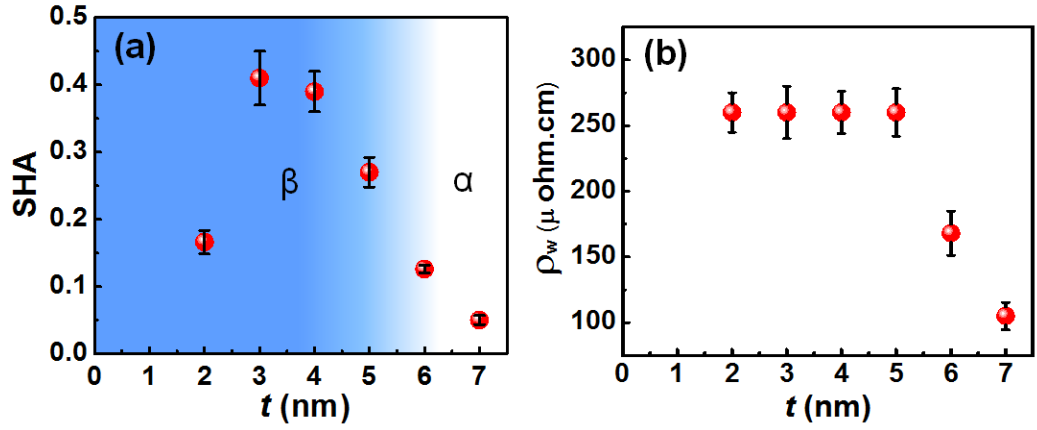

Figure 5. (a) Variation of spin Hall angle with W thickness. Error bars are estimated by considering errors in damping, saturation magnetization and resistivity measurements. The color contrast shows the transition from β to α phase of W. (b) Variation of resistivity of W with thickness ($t$).